\documentclass[%
 reprint,
superscriptaddress,
showpacs,
 amsmath,amssymb,
 aps, prl
]{revtex4-1}

\usepackage{graphicx}
\usepackage{dcolumn}
\usepackage{bm}
\usepackage{hyperref}
\usepackage{subfigure}
\usepackage{comment}
\usepackage{xcolor}
\usepackage[mathlines]{lineno}

\usepackage{ifthen} 
\newboolean{pdflatex}
\setboolean{pdflatex}{true} 

\newboolean{articletitles}
\setboolean{articletitles}{true} 

\newboolean{uprightparticles}
\setboolean{uprightparticles}{false} 

\newboolean{inbibliography}
\setboolean{inbibliography}{false} 

\newboolean{wordcount}
\setboolean{wordcount}{false} 

\ifthenelse{\boolean{wordcount}}%
{\usepackage{bibentry} 
 \usepackage{mciteplus} 
 \usepackage{comment} 
 \excludecomment{acknowledgments} 
 \def\maketitle{} 
}{
}

\graphicspath{{./figs/}}
\DeclareGraphicsExtensions{.pdf,.PDF,png,.PNG}
\usepackage{xspace} 
\makeatletter
\gdef\@ptsize{0} 
\makeatother
\makeatletter
\usepackage{suffix}
\let\latex@section\section
\WithSuffix\def\section*{\secdef\my@section{\latex@section*}}
\def\my@section[#1]#2{}
\makeatother

\newcommand {\bdstlnu} {\overline{B} \rightarrow D^\ast \ell^- \overline{\nu}_\ell}

\newcommand {\barnuell} {\overline{\nu}_\ell}

\newcommand {\ctv} {\cos \theta_V}
\newcommand {\thetav} {\theta_V}
\def\BB      {\ensuremath{B\overline{B}}\xspace}
\def\qsq     {{\ensuremath{q^2}}\xspace}
\def\ctl     {\ensuremath{\cos{\theta_\ell}}\xspace}
\def\ctv     {\ensuremath{\cos{\theta_V}}\xspace}
\def\thetal  {\ensuremath{\theta_\ell}\xspace}
\def\thetav  {\ensuremath{\theta_V}\xspace}
\def\Dstar   {\ensuremath{D^\ast}}

\usepackage{relsize}
\def\babar{\mbox{\slshape B\kern-0.1em{\smaller A}\kern-0.1em B\kern-0.1em{\smaller A\kern-0.2em R}}\;}

\newcommand {\ups} {\Upsilon(4S)}

\newcommand {\eext} {E_{\mbox{\scriptsize extra}}}
\newcommand {\btag} {B_{\mbox{\scriptsize tag}}}

\newcommand {\bsigb} {\overline{B}_{\mbox{\scriptsize sig}}}
\newcommand {\pmiss} {\vec{p}_{\mbox{\scriptsize miss}} }

\newcommand {\Vcb} {|V_{cb}|}
\newcommand {\Vub} {|V_{ub}|}

\def\CP                {{\ensuremath{C\!P}}\xspace}
\def\ellm       {{\ensuremath{\ell^-}}\xspace}
\def\deriv {\ensuremath{\mathrm{d}}}

\newcommand*\patchAmsMathEnvironmentForLineno[1]{%
\expandafter\let\csname old#1\expandafter\endcsname\csname #1\endcsname
\expandafter\let\csname oldend#1\expandafter\endcsname\csname
end#1\endcsname
 \renewenvironment{#1}%
   {\linenomath\csname old#1\endcsname}%
   {\csname oldend#1\endcsname\endlinenomath}%
}
\newcommand*\patchBothAmsMathEnvironmentsForLineno[1]{%
  \patchAmsMathEnvironmentForLineno{#1}%
  \patchAmsMathEnvironmentForLineno{#1*}%
}

\AtBeginDocument{%
\patchBothAmsMathEnvironmentsForLineno{equation}%
\patchBothAmsMathEnvironmentsForLineno{align}%
\patchBothAmsMathEnvironmentsForLineno{flalign}%
\patchBothAmsMathEnvironmentsForLineno{alignat}%
\patchBothAmsMathEnvironmentsForLineno{gather}%
\patchBothAmsMathEnvironmentsForLineno{multline}%
\patchBothAmsMathEnvironmentsForLineno{eqnarray}%
}

\usepackage{xspace} 
\usepackage{upgreek}

\usepackage{relsize}
\def\babar{\mbox{\slshape B\kern-0.1em{\smaller A}\kern-0.1em
    B\kern-0.1em{\smaller A\kern-0.2em R}}\xspace}

\def\MagUp {\mbox{\em Mag\kern -0.05em Up}\xspace}

\ifthenelse{\boolean{uprightparticles}}%
{

 \def\Ppsi        {\ensuremath{\uppsi}\xspace}

 \def\PDelta      {\ensuremath{\Delta}\xspace}                 
 \def\PXi      {\ensuremath{\Xi}\xspace}                 
 \def\PLambda      {\ensuremath{\Lambda}\xspace}                 
 \def\PSigma      {\ensuremath{\Sigma}\xspace}                 
 \def\POmega      {\ensuremath{\Omega}\xspace}                 
 \def\PUpsilon      {\ensuremath{\Upsilon}\xspace}                 
 
 \def\PB      {\ensuremath{\mathrm{B}}\xspace}                 
                  
 \def\PD      {\ensuremath{\mathrm{D}}\xspace}

 \def\PJ      {\ensuremath{\mathrm{J}}\xspace}                 
 \def\PK      {\ensuremath{\mathrm{K}}\xspace}

 \def\Pb      {\ensuremath{\mathrm{b}}\xspace}                 
 \def\Pc      {\ensuremath{\mathrm{c}}\xspace}

 \def\Pi      {\ensuremath{\mathrm{i}}\xspace}

 \def\Pu      {\ensuremath{\mathrm{u}}\xspace}

}
{

 \def\Ppsi        {\ensuremath{\psi}\xspace}                 
                  
 \mathchardef\PDelta="7101
 \mathchardef\PXi="7104
 \mathchardef\PLambda="7103
 \mathchardef\PSigma="7106
 \mathchardef\POmega="710A
 \mathchardef\PUpsilon="7107
                  
 \def\PB      {\ensuremath{B}\xspace}                 
                  
 \def\PD      {\ensuremath{D}\xspace}

 \def\PJ      {\ensuremath{J}\xspace}                 
 \def\PK      {\ensuremath{K}\xspace}

 \def\Pb      {\ensuremath{b}\xspace}                 
 \def\Pc      {\ensuremath{c}\xspace}

 \def\Pi      {\ensuremath{i}\xspace}

 \def\Pu      {\ensuremath{u}\xspace}

}

\makeatletter
\ifcase \@ptsize \relax
  \newcommand{\miniscule}{\@setfontsize\miniscule{4}{5}}
\or
  \newcommand{\miniscule}{\@setfontsize\miniscule{5}{6}}
\or
  \newcommand{\miniscule}{\@setfontsize\miniscule{5}{6}}
\fi
\makeatother

\DeclareRobustCommand{\optbar}[1]{\shortstack{{\miniscule (\rule[.5ex]{1.25em}{.18mm})}
  \\ [-.7ex] $#1$}}

\def\ellm       {{\ensuremath{\ell^-}}\xspace}

\def\uquark    {{\ensuremath{\Pu}}\xspace}

\def\cquark    {{\ensuremath{\Pc}}\xspace}

\def\bquark    {{\ensuremath{\Pb}}\xspace}

  \def\Kbar    {{\kern 0.2em\overline{\kern -0.2em \PK}{}}\xspace}

\def\KorKbar    {\kern 0.18em\optbar{\kern -0.18em K}{}\xspace}

  \def\Dbar    {{\kern 0.2em\overline{\kern -0.2em \PD}{}}\xspace}
\def\D       {{\ensuremath{\PD}}\xspace}

\def\DorDbar    {\kern 0.18em\optbar{\kern -0.18em D}{}\xspace}

\def\Dstar   {{\ensuremath{\D^*}}\xspace}

\def\Bbar    {{\ensuremath{\kern 0.18em\overline{\kern -0.18em \PB}{}}}\xspace}

\def\BorBbar    {\kern 0.18em\optbar{\kern -0.18em B}{}\xspace}

\def\jpsi     {{\ensuremath{{\PJ\mskip -3mu/\mskip -2mu\Ppsi\mskip 2mu}}}\xspace}

  \def\Y#1S{\ensuremath{\PUpsilon{(#1S)}}\xspace}

\def\Lbar        {{\ensuremath{\kern 0.1em\overline{\kern -0.1em\PLambda}}}\xspace}
\def\LorLbar    {\kern 0.18em\optbar{\kern -0.18em \PLambda}{}\xspace}

\def\to                 {\ensuremath{\rightarrow}\xspace}

\def\qsq       {{\ensuremath{q^2}}\xspace}

\def\CP                {{\ensuremath{C\!P}}\xspace}

\def\Vub  {{\ensuremath{V_{\uquark\bquark}}}\xspace}
\def\Vcb  {{\ensuremath{V_{\cquark\bquark}}}\xspace}

\def\AT#1     {\ensuremath{A_{\mathrm{T}}^{#1}}\xspace}           

\def\ctl       {\ensuremath{\cos{\theta_\ell}}\xspace}

\def\C#1      {\ensuremath{\mathcal{C}_{#1}}\xspace}                       
\def\Cp#1     {\ensuremath{\mathcal{C}_{#1}^{'}}\xspace}                    
\def\Ceff#1   {\ensuremath{\mathcal{C}_{#1}^{\mathrm{(eff)}}}\xspace}        
\def\Cpeff#1  {\ensuremath{\mathcal{C}_{#1}^{'\mathrm{(eff)}}}\xspace}       
\def\Ope#1    {\ensuremath{\mathcal{O}_{#1}}\xspace}                       
\def\Opep#1   {\ensuremath{\mathcal{O}_{#1}^{'}}\xspace}                    

\newcommand{\tev}{\ifthenelse{\boolean{inbibliography}}{\ensuremath{~T\kern -0.05em eV}}{\ensuremath{\mathrm{\,Te\kern -0.1em V}}}\xspace}
\newcommand{\gev}{\ensuremath{\mathrm{\,Ge\kern -0.1em V}}\xspace}
\newcommand{\mev}{\ensuremath{\mathrm{\,Me\kern -0.1em V}}\xspace}
\newcommand{\kev}{\ensuremath{\mathrm{\,ke\kern -0.1em V}}\xspace}
\newcommand{\ev}{\ensuremath{\mathrm{\,e\kern -0.1em V}}\xspace}
\newcommand{\gevc}{\ensuremath{{\mathrm{\,Ge\kern -0.1em V\!/}c}}\xspace}
\newcommand{\mevc}{\ensuremath{{\mathrm{\,Me\kern -0.1em V\!/}c}}\xspace}
\newcommand{\gevcc}{\ensuremath{{\mathrm{\,Ge\kern -0.1em V\!/}c^2}}\xspace}
\newcommand{\gevgevcccc}{\ensuremath{{\mathrm{\,Ge\kern -0.1em V^2\!/}c^4}}\xspace}
\newcommand{\mevcc}{\ensuremath{{\mathrm{\,Me\kern -0.1em V\!/}c^2}}\xspace}

\def\deriv {\ensuremath{\mathrm{d}}}

\def\gsim{{~\raise.15em\hbox{$>$}\kern-.85em
          \lower.35em\hbox{$\sim$}~}\xspace}
\def\lsim{{~\raise.15em\hbox{$<$}\kern-.85em
          \lower.35em\hbox{$\sim$}~}\xspace}

\def\tell1  {TELL1\xspace}
\def\ukl1   {UKL1\xspace}

\newcommand{\etal}{\mbox{\itshape et al.}\xspace}

\usepackage{longtable} 
\usepackage{subdepth}

\begin{document}

\preprint{LHCb-PAPER-20XX-YYY}

\onecolumngrid
\small
\begin{flushleft}
\babar-PUB-19/001\\
SLAC-PUB-17420
\end{flushleft}
\normalsize

\title{Extraction of form factors from a four-dimensional angular analysis of $\overline{B} \rightarrow D^\ast \ell^- \overline{\nu}_\ell$}

\author{J.~P.~Lees}
\author{V.~Poireau}
\author{V.~Tisserand}
\affiliation{Laboratoire d'Annecy-le-Vieux de Physique des Particules (LAPP), Universit\'e de Savoie, CNRS/IN2P3,  F-74941 Annecy-Le-Vieux, France}
\author{E.~Grauges}
\affiliation{Universitat de Barcelona, Facultat de Fisica, Departament ECM, E-08028 Barcelona, Spain }
\author{A.~Palano}
\affiliation{INFN Sezione di Bari and Dipartimento di Fisica, Universit\`a di Bari, I-70126 Bari, Italy }
\author{G.~Eigen}
\affiliation{University of Bergen, Institute of Physics, N-5007 Bergen, Norway }
\author{D.~N.~Brown}
\author{Yu.~G.~Kolomensky}
\affiliation{Lawrence Berkeley National Laboratory and University of California, Berkeley, California 94720, USA }
\author{M.~Fritsch}
\author{H.~Koch}
\author{T.~Schroeder}
\affiliation{Ruhr Universit\"at Bochum, Institut f\"ur Experimentalphysik 1, D-44780 Bochum, Germany }
% ...
\collaboration{The \babar\ Collaboration}
\noaffiliation



\begin{abstract}
\noindent An angular analysis of the decay ${\overline{B} \rightarrow D^\ast \ell^- \overline{\nu}_\ell}$, $\ell\in\{e,\mu\}$, is reported using the full $e^+e^-$ collision data set collected by the \babar experiment at the $\Upsilon(4S)$ resonance. One $B$ meson from the ${\Upsilon(4S)\to B\overline{B}}$ decay is fully reconstructed in a hadronic decay mode, which constrains the kinematics and provides a determination of the neutrino momentum vector. The kinematics of the semileptonic decay is described by the di-lepton mass squared, $q^2$, and three angles. The first unbinned fit to the full four-dimensional decay rate in the Standard Model is performed in the so-called Boyd-Grinstein-Lebed approach, which employs a generic $q^2$ parameterization of the underlying form factors based on crossing symmetry, analyticity and QCD dispersion relations for the amplitudes. A fit using the more model-dependent Caprini-Lellouch-Neubert (CLN) approach is performed as well. Our form factor shapes show deviations from previous fits based on the CLN parameterization. The latest form factors also provide an updated prediction for the branching fraction ratio $\mathcal{R}(D^\ast)\equiv \mathcal{B}(\overline{B}\to D^\ast \tau^- \bar{\nu}_\tau)/\mathcal{B}({\overline{B} \rightarrow D^\ast \ell^- \overline{\nu}_\ell})=0.253 \pm 0.005$. 
Finally, using the well measured branching fraction for the ${\overline{B} \rightarrow D^\ast \ell^- \overline{\nu}_\ell}$ decay, a value of $|V_{cb}|=(38.36\pm 0.90)\times10^{-3}$ is obtained that is consistent with the current world average for exclusive ${\overline{B}\to D^{(\ast)}\ell^- \overline{\nu}_\ell}$ decays and remains in tension with the determination from inclusive semileptonic $B$ decays to final states with charm.

\end{abstract}

\let\oldmaketitle\maketitle
\renewcommand\maketitle{{\bfseries\boldmath\oldmaketitle}}

\maketitle

The Cabibbo-Kobayashi-Maskawa (CKM) matrix~\cite{Cabibbo:1963yz,Kobayashi:1973fv}, $V_{\scriptsize {\rm CKM}}$, describing quark flavor mixing due to the charged weak current, is one of pillars of the Standard Model (SM) of particle physics. It contains the only source of charge-parity (\CP) violation in the SM. Validating this picture requires precise determinations of the CKM matrix elements $|\Vub|$ and $|\Vcb|$. These are measured by the tree-level semileptonic decays, $b\to \{u,c\}\ellm \barnuell$, where $\ell$ refers to an electron or muon. The hadronization of the final-state $\{u,c\}$ quark can be probed via inclusive or exclusive final states, the theoretical treatment being quite different for the two processes. For the heavy-to-heavy $b\to c$ transition, the inclusive and exclusive procedures use an operator product expansion and form factors based on heavy quark effective theory (HQET), respectively~\cite{Bigi:2017jbd}. The theoretical and experimental uncertainties are different in the two cases, and a long-standing tension of about $3\sigma$ exists between them, with the inclusive results systematically higher than the exclusive ones, for both $|\Vub|$ and $|\Vcb|$. The different results from inclusive and exclusive measurements could arise from non-SM physics. This motivates better quantification of uncertainties in the measurements and underlying theoretical treatment of strong interaction effects. Recently, several authors have pointed out~\cite{Grinstein:2017nlq,Bigi:2017njr,Bigi:2017jbd}, based on fits to unpublished Belle data~\cite{Abdesselam:2017kjf}, that removing HQET constraints in the theoretical parameterization of the $\overline{B}\to D^\ast$ form factors can reduce the tension between inclusive and exclusive $|\Vcb|$ determinations. The measurement described here is a test of this suggestion.

The $\bdstlnu$~\footnote{The inclusion of charge-conjugate decay modes is implied throughout this Letter} process, with the subsequent ${D^\ast \to D \pi}$ decay, requires four independent kinematic variables to fully parametrize the final state. For the analysis presented in this Letter, we adopt the customary choice~\cite{spd-paper} of the di-lepton invariant mass squared, $\qsq$, the helicity angles of the $D$ and $\ellm$, $\thetav$ and $\thetal$, respectively, and the angle $\chi$ between the hadronic and leptonic two-body decay planes. Denoting $\deriv\Omega = \deriv\!\ctl\,\deriv\!\ctv\,\deriv\chi$, the four-dimensional differential rate assuming massless leptons in the SM is~\cite{spd-paper}
\begin{align}
\label{eqn:full_angular_distribution_V}
& \frac{\deriv\Gamma }{\deriv \qsq \deriv\Omega} = \bigg[\left(H_+^2  (1 -\cos\thetal)^2 +   H_-^2 (1 +\cos\thetal)^2\right)\sin^2\thetav + \bigg. \nonumber  \\ 
&  2 H_0 \sin\thetal \sin2\thetav \cos \chi \left( H_+ (1-\cos\thetal) - H_- (1+\cos\thetal)\right) \nonumber \\
& \bigg. + 4 H_0^2 \sin^2\thetal \cos^2\thetav - 2 H_+H_- \sin^2\thetal \sin^2\thetav\cos2\chi  \bigg] \nonumber \\
&\hspace{1.5cm} \times \frac{3}{8(4\pi)^4} G_F^2\eta^2_{\rm EW} |V_{cb}|^2 \frac{k  \qsq}{m^2_B} \mathcal{B}(D^\ast \to D\pi),
\end{align}
where $k = \sqrt{ (m^2_B-q^2 +m_{D^\ast}^2)^2/4m^2_B -m^2_{D^\ast}}$ is the $D^\ast$ momentum in the $B$ rest frame, $\eta_{\rm EW}=1.0066$~\cite{Sirlin:1981ie,Grinstein:2017nlq} denotes leading electroweak corrections, and $G_F$ is the Fermi decay constant. In the SM, the helicity amplitudes $H_{\pm,0}$ are the real functions
\begin{align} 
H_0 &= \frac{1}{2 m_{D^\ast} \sqrt{q^2}} \Big((m^2_B - m^2_{D^\ast} - q^2) (m_B +m_{D^\ast} ) A_1(\qsq)\Big. \nonumber  \\
              & \hspace{2.5cm} \Big. - \frac{4 m^2_B k^2}{m_B +m_{D^\ast}}  A_2(\qsq) \Big), \\
H_{\pm} &= (m_B + m_{D^\ast}) A_1(q^2) \mp \frac{2 m_B k}{(m_B + m_{D^\ast})} V(\qsq),
\label{eqn:hel_amps}
\end{align}
expressed here in terms of the conventional axial-vector and vector form factors, $\{A_1, A_2, V\}$, as in Caprini {\em et al.} (CLN)~\cite{Caprini:1997mu}. In the Boyd {\em et al.} (BGL)~\cite{Boyd:1995sq} approach, the form factors are written as ${f = (m_B+m_{\Dstar})A_1}$, ${F_1 = \sqrt{\qsq} H_0}$ and ${g = 2V/ (m_B +m_{\Dstar})}$. The BGL formalism parameterizes the $i^{th}$ form factor, $F_i$, in the most generic form, based on crossing symmetry, analyticity and QCD dispersion relations, as
\begin{align}
\label{eqn:bgl_ff_expansion}
F_i(z) = \displaystyle \frac{1}{P_i(z) \phi_i(z)} \sum_{n=0}^N a^i_n z^n.
\end{align}
The expansion parameter $z$ is given by
\begin{align}
\label{eqn:bgl_z}
z(t,t_0) &= \displaystyle \frac{\sqrt{t_+ - t} - \sqrt{t_+ - t_0} }{\sqrt{t_+ - t} + \sqrt{t_+ - t_0}},
\end{align}
and is small in the physical region. Here $t\equiv \qsq$, $t_\pm \equiv (m_B \pm m_{\Dstar})^2$ and $t_0= t_+ - \sqrt{t_+(t_+-t_-)}$. We adopt the Blaschke factors, $P_i(z)$, corresponding to removal of the $B_c$ poles of the $BD^\ast$ system, and the outer functions, $\phi_i(z)$, from Refs.~\cite{Boyd:1997kz,Bigi:2017jbd}. The BGL coefficients in Eq.~\ref{eqn:bgl_ff_expansion} satisfy the relations $\sum_n |a^i_n|^2\leq 1$, known as unitarity constraints. The CLN~\cite{Caprini:1997mu} formalism makes similar expansions up to cubic terms, but imposes heavy-quark symmetry relations and QCD sum rules to relate the expansion parameters. The resultant forms are expressed in terms of a reduced set of a slope, $\rho^2_{D^\ast}$, and two normalization parameters, $R_{1,2}(1)$. 

In this Letter, employing a data sample of $471\times10^6$ \BB pairs~\cite{Lees:2013rw} produced at the $\ups$ resonance and collected by the \babar detector~\cite{Aubert:2001tu,TheBABAR:2013jta}, a full four-dimensional analysis of the $\bdstlnu$ decay rate corresponding to Eq.~\ref{eqn:full_angular_distribution_V} is reported. One of the $B$ mesons, referred to as the tag-side $B$, is fully reconstructed via hadronic decays, allowing for the missing neutrino $4$-momentum, $p_{\rm miss}$, to be explicitly reconstructed on the signal-side $B$, since the initial $e^\pm$ $4$-momenta are known. The hadronic tagging algorithm uses charm-meson seeds ($D^{(\ast)}$, \jpsi) combined with ancillary charmless light hadrons ($\pi/K$), and is the same as in several previous \babar analyses~\cite{Aubert:2001tu, TheBaBar:2016xwe, Lees:2015eya}. From the remaining particles in the event after the tag-$B$ reconstruction, a $D^0$ meson reconstructed via one its three cleanest decay modes, $K^-\pi^+$, $K^-\pi^-\pi^0$ or $K^-\pi^+\pi^-\pi^+$, is combined with a $\pi^0$ or $\pi^+$, to form a $D^{\ast 0}$ or $D^{\ast+}$, respectively. For each $D^\ast$ candidate, the reconstructed invariant mass of the $D^0$ and the difference of the reconstructed masses, $\Delta m \equiv (m_{D^\ast} - m_D)$, are required to be within four standard deviations of the expected resolution from their nominal values, at this stage. The $D^\ast$ is combined with a charged lepton $\ell\in\{e,\mu\}$, with the laboratory momentum of the lepton required to be greater than 0.2~GeV and 0.3~GeV for $e$ and $\mu$, respectively. The six $D^\ast$ decay modes along with the two charged lepton species comprise twelve signal channels that are processed as independent data samples. No additional tracks are allowed in the event. The entire event topology, $e^+e^- \to \ups\to \btag\bsigb(\to D^\ast \ellm \barnuell)$ is considered in a kinematic fit including constraints on the beam spot, relevant secondary decay vertices and masses of the reconstructed $\btag$, $\bsigb$, $D^{(\ast)}$ and the missing neutrino. The $\chi^2$-probability from this highly constrained fit is used as the main discriminant against background. To reject candidates with additional neutral energy deposits, $\eext$ is defined as the sum of the energies of the good quality photons not utilized in the event reconstruction. The variable $\eext$ is required to be less than 0.4~GeV to 0.6~GeV, depending on the $D^{(\ast)}$ modes. Only candidates satisfying $\qsq\in [0.2,10.2]$~GeV$^2$ are retained. In events with multiple selected candidates, only the candidate with the highest $\chi^2$-probability from the kinematic fit is retained. 

\begin{figure}
\centering
\includegraphics[width=0.49\linewidth]{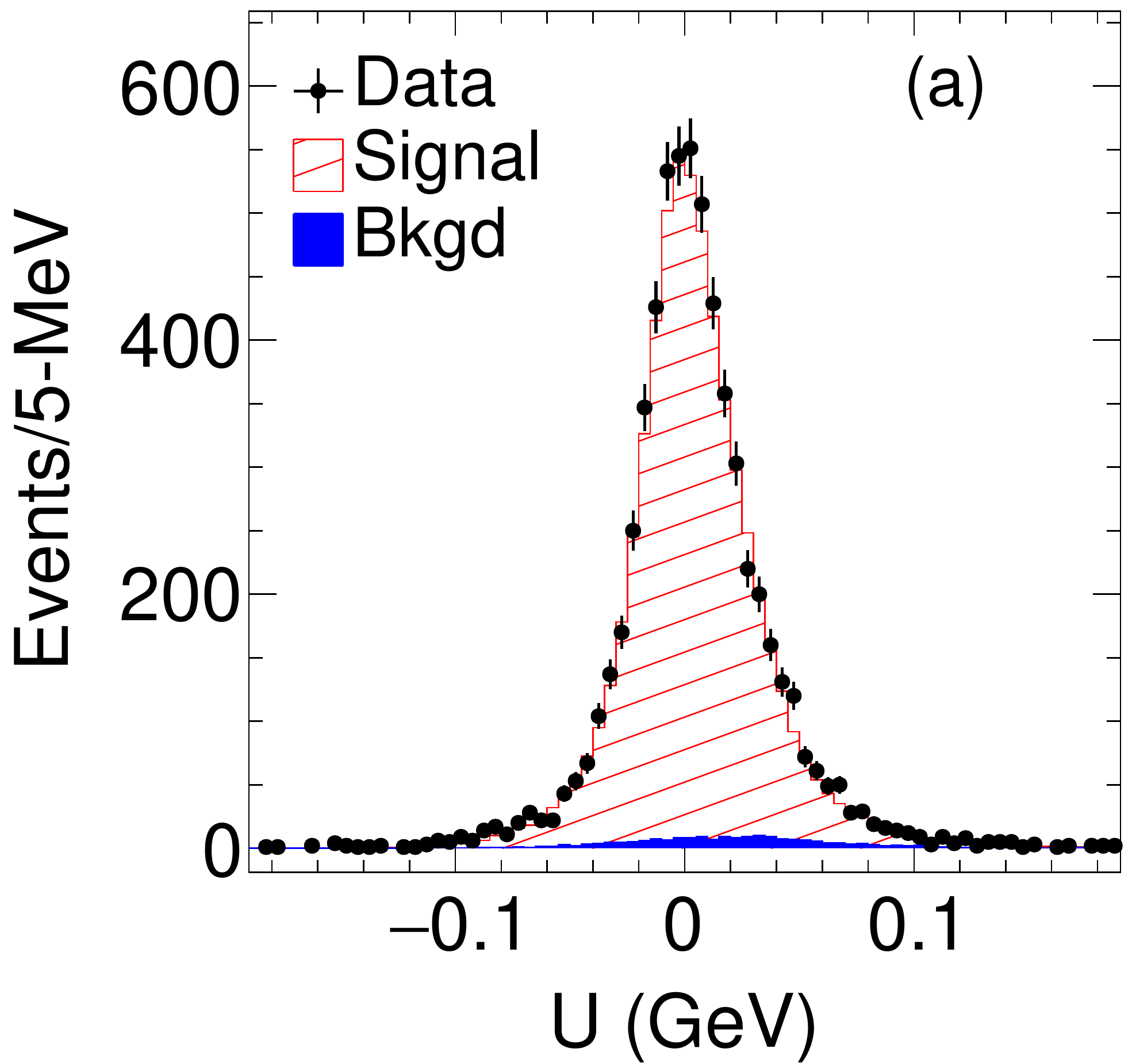}
\includegraphics[width=0.49\linewidth]{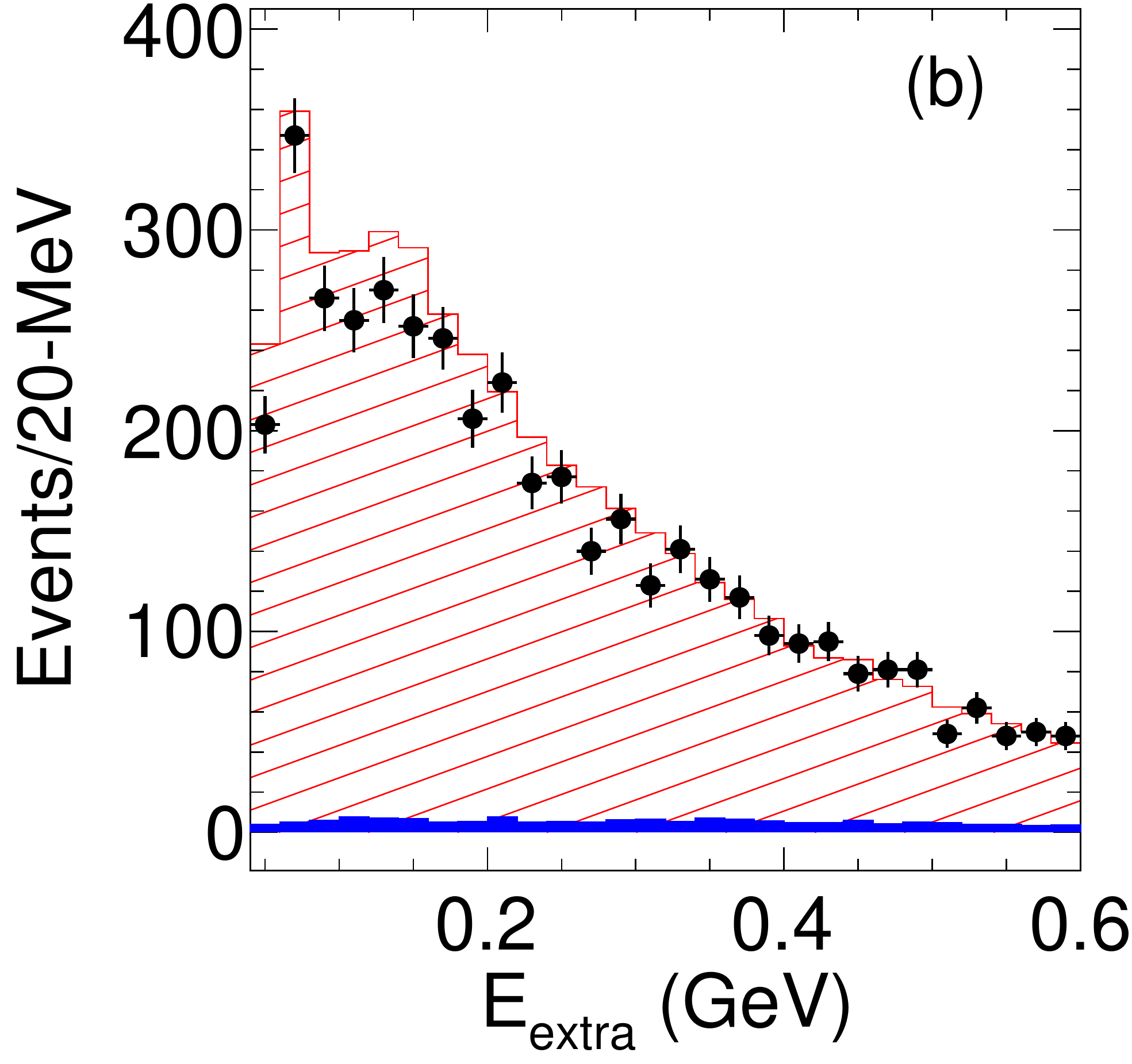}
\caption[]{\label{fig:data_mc_U} Comparisons between data and generic $\BB$ simulation in the discriminating variables (a) $U$ and (b) $\eext$. For each plot, selections in all other variables have been applied.}
\end{figure}

After all selections, the overall background level is estimated to be $\sim 2\%$, using a simulation of generic ${\ups\to \BB}$ events, where both $B$ mesons decay to any allowed final state. All selected events enter the four-dimensional angular fit; the small remnant background is treated as a source of systematic uncertainty. Figure~\ref{fig:data_mc_U}a shows the comparison between data and simulation in the variable $U= E_{\mbox{\scriptsize miss}} - |\pmiss|$, where the resolution in the neutrino reconstruction has been weighted in the signal part of this simulation to match that in the data. Here $E_{\mbox{\scriptsize miss}}$ and $\pmiss$ correspond to the missing neutrino energy and momentum, respectively. Figure~\ref{fig:data_mc_U}b shows the comparison in the discriminating variable $\eext$. The efficiency in $\eext$ in the $\eext\to 0$ signal region does not affect the angular analysis, so that an exact agreement is not required. The generic $\BB$ simulation agrees with the data in all kinematic-variable distributions in the sideband regions, validating its use to estimate the background in the signal region. The final requirement is $|U|\leq 90$~MeV. The total number of selected candidates at this stage is 6112, with the estimated signal yield being around 5932.

In addition to the generic $\BB$ simulation sample used for the data analysis where both $B$-mesons are decayed generically, a separate category of $\BB$ simulation is employed where the $\btag$ is decayed generically, but ${\bsigb\to D^\ast(\to D\pi)\ellm\barnuell}$ is decayed uniformly in $\deriv \qsq d\Omega$ at the generator level. This latter sample is used to correct for detector acceptance effects in the fit to Eq.~\ref{eqn:full_angular_distribution_V} employing numeric computation of the normalization integrals as described in Ref.~\cite{Chung}. The simulation undergoes the same reconstruction and selection steps as the data sample. The uniformly generated simulation weighted by the fit results match the data in all distributions, as discussed later.

Unbinned maximum-likelihood fits to the \babar data are performed employing the four-dimensional decay rate given by Eq.~\ref{eqn:full_angular_distribution_V}. The likelihood calculation treats all events in the data sample as signal and the small residual background is accounted for by subtracting from the log likelihood a contribution estimated from generic $\BB$ simulation. The fits are performed in two variants, for each of the BGL and CLN parameterizations. For the nominal \babar-only variant, the negative log likelihood (NLL) is of the non-extended type, implying that the overall normalization factor is not imposed. This fit is used to extract the three form factors in a fashion insulated from systematic uncertainties related to the normalization, in particular with the estimation of the $\btag$ yield. 

To extract $|\Vcb|$, a second version of the fit is performed, where the integrated rate $\Gamma$ is converted to a branching fraction, $\mathcal{B}$, as $\Gamma = \mathcal{B} /\tau_B$, where $\tau_B$ is the $B$-meson lifetime. The latest HFLAV~\cite{HFLAV16} values of $\mathcal{B}$ and $\tau_B$, for $B^0$ and $B^-$ mesons, are employed as additional Gaussian constraints to the \babar-only NLL, and the entire fit is repeated. Two other constraints are employed. First, a lattice calculation from the Fermilab Lattice and MILC collaborations~\cite{Bailey:2014tva} gives the value of $h_{A1}(1) = (m_B+ m_{\Dstar}) A_1(q^2_{\rm max})/(2\sqrt{m_B m_{\Dstar}})$ at the zero recoil point, $q^2_{\rm max}\equiv(m_B- m_{\Dstar})^2$. Second, at the zero recoil point, the relation $F_1(q^2_{\rm max}) = (m_B - m_\Dstar) f(q^2_{\rm max})$ is used to express $a_0^{F_1}$ in terms of the remaining BGL coefficients in $f$ and $F_1$. Therefore, $a_0^{F_1}$ is not a free parameter in the fit, but is derived from the remaining parameters. The small isospin dependence of these constraints, arising from the differences $m_{B^+} - m_{B^0}$ and $m_{D^{\ast 0}} - m_{D^{\ast +}}$, is ignored in the calculation. 

Given the statistical power of our data, we truncate the BGL expansion at $N=1$ to avoid the violation of unitarity constraints due to poorly determined parameters. To ensure that a global minimum for the NLL is reached, 1000 instances of the BGL fits are executed, with uniform sampling on [-1,+1] for the starting values of the $a_n$ coefficients. Among convergent fits, a unique minimum NLL is always found, up to small variations in the least significant digits in the fit parameters.

Many sources of systematic uncertainties cancel in this analysis, since no normalization is required from the \babar data sample. Tracking efficiences in simulation show no significant dependence on $\qsq$ or $\{\ctl,\ctv,\chi\}$. To account for the resolutions in the reconstructed kinematic variables, the normalization of the probablity density function in the fit is performed using reconstructed variables from the simulation. The dominant systematic uncertainty comes from the remnant background that can pollute the angular distributions. To estimate its effect on the fit results, the fit procedure is repeated excluding the background subtraction and the difference in the results is taken as the systematic uncertainty. 

\begin{table}
\begin{center}
\begin{tabular}{c|c|c|c|c|c} \hline \hline
$a^f_0\times 10^2$ & $a^f_1\times 10^2$ & $a^{F_1}_1\times 10^2$ & $a^g_0\times 10^2$ & $a^g_1\times 10^2$ & $|\Vcb|\times 10^3$ \\ \hline
$1.29$ & $1.63$ & $0.03$ & $2.74$ & $8.33$ & $38.36$ \\
$\pm 0.03 $ & $\pm 1.00 $ & $\pm 0.11 $ & $\pm 0.11 $ & $\pm 6.67 $ & $\pm 0.90$\\\hline\hline
\end{tabular}
\caption{The $N=1$ BGL expansion results of this analysis, including systematic uncertainties.}
\label{table:bgl_n=1}
\end{center}
\end{table}

\begin{table}
\begin{center}
\begin{tabular}{c|c|c|c} \hline \hline
$\rho^2_{D^\ast}$ & $R_1(1)$ & $R_2(1)$ & $|\Vcb|\times 10^3$ \\ \hline
$0.96\pm 0.08$ & $1.29\pm 0.04$ & $0.99\pm 0.04$ & $38.40\pm 0.84$ \\ \hline\hline
\end{tabular}
\caption{The CLN fit results from this analysis, including systematic uncertainties.}
\label{table:cln}
\end{center}
\end{table}

Table~\ref{table:bgl_n=1} summarizes the main results from the BGL fits, including $|\Vcb|$. Several checks are performed to ensure stability of the results. Cross-checks are performed via separate fits to the $B^0$ and $B^-$ isospin modes that have charged and neutral pions for the soft pion in $D^\ast \to D\pi$~\cite{deBoer:2018ipi}. Cross-checks are also performed for separate fits to the two lepton species. Results are found to be compatible within the statistical uncertainties and thus no additional uncertainty is quoted from these checks. The values of $|\Vcb|\times 10^3$, including only statistical uncertainties, for the $e$, $\mu$, $B^0$, $B^-$ separated fits are $38.59\pm 1.15 $, $38.24\pm 1.05 $, $38.03\pm1.05$ and $38.68\pm1.16$, respectively. The use of $t_0=t_-$ in the BGL expansion, as in Refs.~\cite{Grinstein:2017nlq,Bigi:2017njr,Bigi:2017jbd} also gives results consistent with Table~\ref{table:bgl_n=1}. Table~\ref{table:cln} reports the corresponding results from the CLN fits. The value of $|\Vcb|$ is consistent between the BGL and CLN based fits.

\begin{figure*}
\begin{center}
\includegraphics[width=\linewidth]{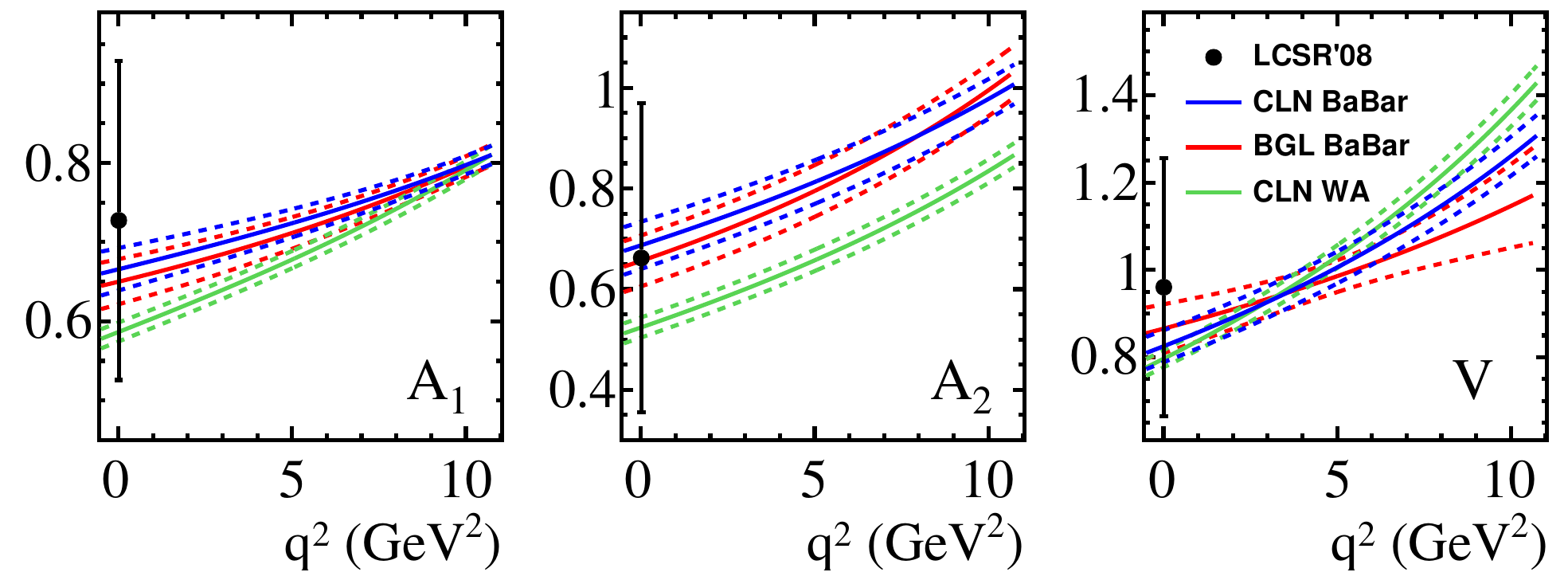}
 \caption{Comparison between the \babar BGL/CLN and CLN-WA~\cite{HFLAV16} form factors, $\{A_1,A_2,V\}$. Also shown is the LCSR prediction at $\qsq=0$~\cite{Faller:2008tr}. The error bands are depicted by the dashed curves and include both statistical and systematic uncertainties.}
 \label{fig:a1a2v_ichep}
\end{center}
\end{figure*}

Figure~\ref{fig:a1a2v_ichep} shows the comparisons of the \babar BGL/CLN results with the CLN world average (CLN-WA)~\cite{HFLAV16} as well as light-cone sum rules (LCSR) at the maximum recoil from Ref.~\cite{Faller:2008tr}. Phenomenologically, the most important feature in Fig.~\ref{fig:a1a2v_ichep} is the discrepancy between CLN-WA and the \babar fits, while within \babar, both CLN and BGL parameterizations yield comparable results. Numerically, the $p$-value of the consistency check in the three CLN fit parameters, between CLN-\babar and CLN-WA is 0.0017. 

For $|\Vcb|$, the result obtained here is well below the value determined from inclusive decays. This is in contrast with results from several recent analyses using the BGL parameterization based on unpublished Belle data~\cite{Abdesselam:2018nnh,Abdesselam:2017kjf,Grinstein:2017nlq,Bigi:2017njr,Bigi:2017jbd}, where larger values, close to the inclusive result, were typically obtained. 

\begin{figure*}
\begin{center}
\includegraphics[width=\linewidth]{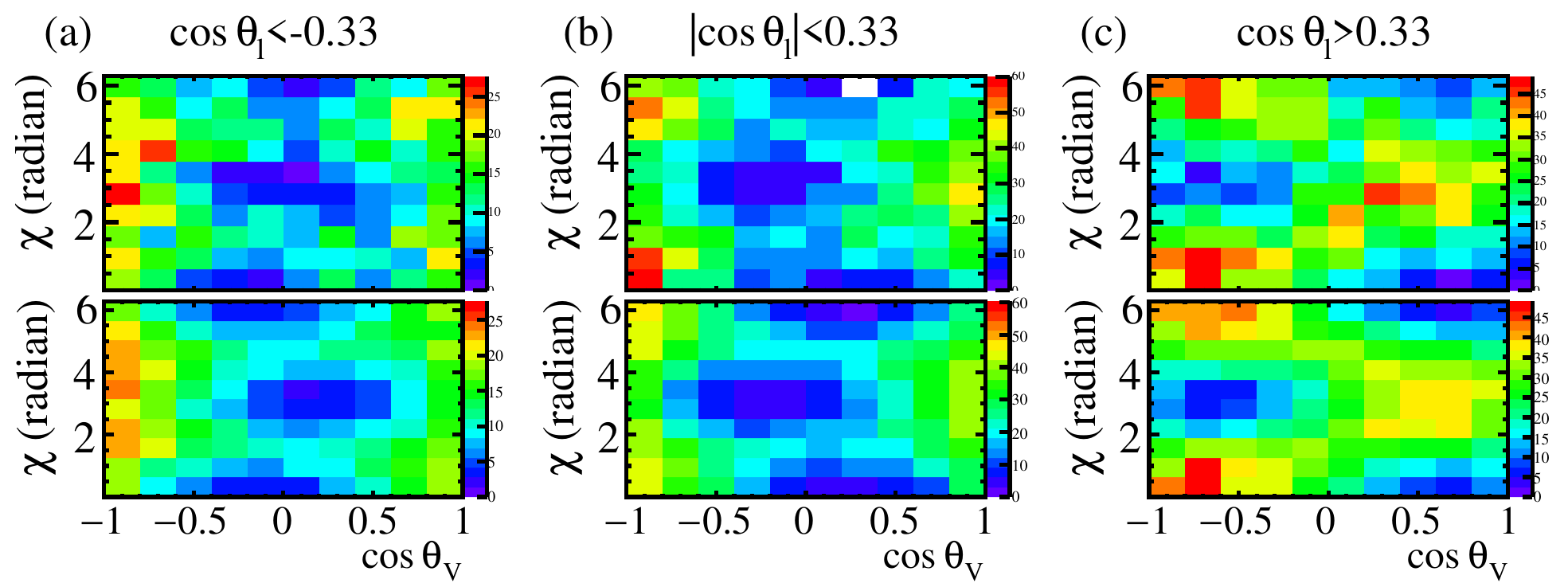}
 \caption{Comparisons as binned scatter plots between the \babar data (top row) and simulation weighted by the BGL fit result (bottom row) in (a) backward, (b) mid and (c) forward angles in \ctl. The multidimensional features in the data are well represented by the model. The $z$-axes indicate the number of events in each bin and the simulation is normalized to the number of data events. \label{fig:2d_data_mc_bgl}}
\end{center}
\end{figure*}

Figure~\ref{fig:2d_data_mc_bgl} shows the two-dimensional scatter plots in $\ctv$ and $\chi$ in three bins of \ctl and integrated over the \qsq spectrum, between the data (top row) and simulation (bottom row) after acceptance and reconstruction effects, weighted by the results of the BGL fit. The binned $\chi^2$ differences between the data and weighted simulation referring to Fig.~\ref{fig:2d_data_mc_bgl} are (a) 103, (b) 89 and (c) 96, evaluated over 100 bins. The corresponding values for the four one-dimensional projections evaluated over 20 bins are 22, 23, 26 and 18, for $\qsq$, $\ctl$, $\ctv$ and $\chi$, respectively. Within uncertainties, the weighted simulation consistently matches the data. 

The differential rate in Eq.~\ref{eqn:full_angular_distribution_V} holds under the assumption that the outgoing charged lepton is massless, a valid approxmimation for $\ell\in\{e,\mu\}$. For the $\tau$ lepton, additional terms appear in the differential rate, $\Gamma(\qsq,m_\ell)$, depending on the lepton mass~\cite{Bigi:2017jbd}. The BGL form factors reported in this Letter lead to an updated prediction for
\begin{align}
\mathcal{R}(D^\ast) \equiv \frac{  \int_{m^2_\tau}^{q^2_{\rm max}} \Gamma(\qsq,m_\tau) d\qsq}{ \int_{m^2_\ell}^{q^2_{\rm max}} \Gamma(\qsq,m_\ell) d\qsq  },
\end{align}
where $\ell=\{e,\mu\}$. An $N=1$ BGL expansion for the additional scalar form factor is performed following Gambino \etal~\cite{Bigi:2017jbd}, using the HQET prediction at zero recoil, with a conservative estimate for the uncertainty. At maximum recoil, instead of employing the LCSR form factors~\cite{Faller:2008tr} with large uncertainties that were adopted in Ref.~\cite{Bigi:2017jbd}, the present \babar result is employed. These values at the two ends of the $\qsq$ spectra completely specify the scalar form factor in the linear expansion. The resultant SM prediction is
\begin{align}
\label{eqn:babar_rdst}
\mathcal{R}(D^\ast)\Large|^{\rm {\scriptsize SM}}_{\rm {\scriptsize \babar}} = 0.253 \pm 0.005.
\end{align}
For a different choice of $t_0=t_-$, a value $0.253 \pm 0.005$ is found, consistent with the above. The result is consistent with the CLN based calculation of $0.252 \pm 0.003$ in Ref.~\cite{Fajfer:2012vx}, although with a larger uncertainty, mostly driven by the uncertainty in the scalar form factor at zero recoil, from HQET~\cite{Bigi:2017jbd}. The degree of HQET violation is an important consideration, impacting the uncertainties, although the central value of $\mathcal{R}(D^\ast)$ is largely unaffected. It is important to note that the experimental measurement of $\mathcal{R}(D^\ast)$ might be sensitive to variations in the BGL form factors since the overall efficiency calculation for the measurement is a convolution of the form factor model and the four-dimensional detector acceptance function.

In summary, using the \babar \BB data sample with one of the $B$ mesons fully reconstructed in hadronic modes, an unbinned four-dimensional fit to tagged ${\bdstlnu}$ decays is performed to extract the form factors in the more model-independent formalism of BGL as well as the model-dependent CLN method. The \babar form factors show differences with CLN-WA. The value of $|\Vcb|$ is found to be lower than those obtained in recent BGL analyses based on unpublished Belle data~\cite{Abdesselam:2018nnh,Abdesselam:2017kjf,Grinstein:2017nlq,Bigi:2017njr,Bigi:2017jbd} that did not use a four-dimensional fit~\footnote{Since the current results were posted on the {\tt arXiv}, an updated version of Ref.~\cite{Abdesselam:2018nnh} appeared in which the $|\Vcb|$ values determined from the BGL and CLN fits are nearly identical.} The tension with inclusive determinations of $|\Vcb|$ persists, even with the more model-independent BGL parameterization of the form factors. The central value of the SM $\mathcal{R}(D^\ast)$ prediction based on a BGL parameterization is consistent with the previous CLN based prediction of Ref.~\cite{Fajfer:2012vx}, but with a larger uncertainty, thereby reducing the overall tension with the latest average of experimental measurements.  An extended version of the results presented here, including unfolded four-dimensional angular moments will be presented in a forthcoming publication.~\footnote{The numerical data presented here can be availed in ASCII format in the zipped source file of the {\tt arXiv} submission.}

\begin{acknowledgments}
\section*{Acknowledgements}
We are grateful for the excellent luminosity and machine conditions provided by our PEP-II colleagues, and for the substantial dedicated effort from the computing organizations that support \babar. The collaborating institutions wish to thank SLAC for its support and kind hospitality. This work is supported by DOE and NSF (USA), NSERC (Canada), CEA and CNRS-IN2P3 (France), BMBF and DFG (Germany), INFN (Italy), FOM (The Netherlands), NFR (Norway), MSHE (Russia), MINECO (Spain), STFC (United Kingdom), BSF (USA-Israel). Individuals have received support from the Marie Curie EIF (European Union) and the A. P. Sloan Foundation (USA).

\end{acknowledgments}

\ifthenelse{\boolean{wordcount}}{ 
  \bibliographystyle{plainat}
  \nobibliography{main-prl}
}{
 \bibliography{main-prl}
}

\ifthenelse{\boolean{wordcount}}{}{%
  \clearpage
  \appendix

}

\end{document}